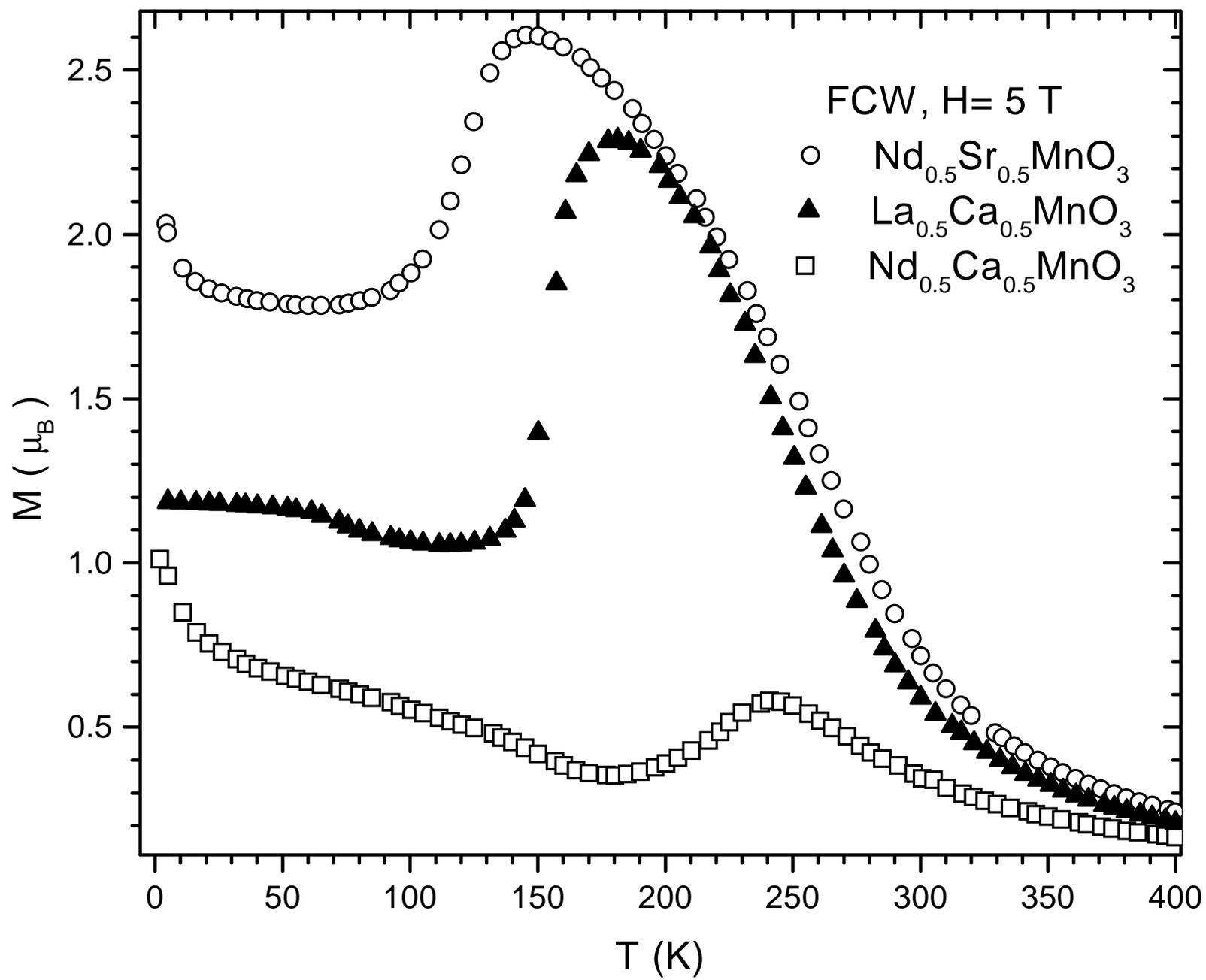

Figure 1

# Study of magnetic and specific heat measurements at low temperatures in Nd$_{0.5}$Sr$_{0.5}$MnO$_3$ and Nd$_{0.5}$Ca$_{0.5}$MnO$_3$

J. López[a], P. N. Lisboa-Filho[a], O. F. de Lima[b] and F. M. Araujo-Moreira[a]

[a]D. de Física, U. Federal de São Carlos, C.P. 676, São Carlos, SP, 13565-905, Brazil.

[b]Instituto de Física Gleb Wataghin, UNICAMP, 13083-970, Campinas, SP, Brazil

**Abstract**

The magnetization at low temperatures for Nd$_{0.5}$Sr$_{0.5}$MnO$_3$ and Nd$_{0.5}$Ca$_{0.5}$MnO$_3$ samples showed a rapid increase with decreasing temperatures, contrary to a La$_{0.5}$Ca$_{0.5}$MnO$_3$ sample. Specific heat measurement at low temperatures showed a Schottky-like anomaly for the first two samples. However, there is not a straight forward correlation between the intrinsic magnetic moment of the Nd$^{3+}$ ions and the Schottky-like anomaly.

**Corresponding autor**: Juan López Linares

**Postal Address**: D. de Física, U. Federal de São Carlos, Caixa Postal - 676, São Carlos, SP, 13565-905, BRAZIL

**Fax**: 0055 16 261 4835

**E-mail**: jlopez@df.ufscar.br

Charge ordering compounds have a large variation of resistivity and magnetization as a function of temperature and magnetic field[1,2]. In two previous reports[3,4] we showed that polycrystalline samples of $La_{0.5}Ca_{0.5}MnO_3$ and $Nd_{0.5}Sr_{0.5}MnO_3$ also presented an unusual magnetic relaxation behavior after applying and turning off a 5 T magnetic field. Here, we briefly report specific heat measurements at low temperatures and zero magnetic field for polycrystalline $Nd_{0.5}Sr_{0.5}MnO_3$ and $Nd_{0.5}Ca_{0.5}MnO_3$ compounds.

Polycrystalline samples of $La_{0.5}Ca_{0.5}MnO_3$ were prepared by the conventional solid-state method described elsewhere[4]. On the other hand, polycrystalline samples of $Nd_{0.5}Sr_{0.5}MnO_3$ and $Nd_{0.5}Ca_{0.5}MnO_3$ were prepared by a sol-gel method[4]. X-ray diffraction measurements indicated high quality samples in all cases. Magnetization measurements were performed with a standard MPMS-5T SQUID magnetometer from Quantum Design. Specific heat measurements were done with a commercial PPMS calorimeter from Quantum Design.

Figure 1 shows the magnetization temperature dependence of the three studied samples with an applied magnetic field of 5 T. Measurements were done with increasing temperatures, after applied the magnetic field at 2 K, increased the temperature up to 400 K and decreased it again to 2 K (field cooling warming, or FCW). All compounds showed an antiferromagnetic transition around $T_N=160$ K and presented charge and orbital ordering at low temperatures.

The transition temperature to the charge ordered phase ($T_{CO}$) is approximately equal to $T_N$ for $La_{0.5}Ca_{0.5}MnO_3$ and $Nd_{0.5}Sr_{0.5}MnO_3$, but $T_{CO}$ is much higher than $T_N$ for $Nd_{0.5}Ca_{0.5}MnO_3$. The peak at 250 K in figure 1 for $Nd_{0.5}Ca_{0.5}MnO_3$ indicates the charge ordering transition. Another intriguing feature in the $Nd_{0.5}Ca_{0.5}MnO_3$ data is the absence of a maximum at $T_N$, contrary to the two other compounds. These results have been confirmed with X-rays and neutron diffraction studies[5,6]. Figure 1 also indicates a ferromagnetic transition about 230 and 250 K for $La_{0.5}Ca_{0.5}MnO_3$ and $Nd_{0.5}Sr_{0.5}MnO_3$ samples, respectively.

Furthermore, the magnetization at low temperatures showed a rapid increase with decreasing temperatures for $Nd_{0.5}Sr_{0.5}MnO_3$ and $Nd_{0.5}Ca_{0.5}MnO_3$. This is in contrast with the $La_{0.5}Ca_{0.5}MnO_3$ sample, where the value of magnetization presented a plateau for the same temperature interval. The reason of this increase in magnetization at low temperatures is based on the intrinsic magnetic moment of $Nd^{3+}$ ions, contrary to $La^{3+}$ ions where this value is zero. However, as was found in the neutron diffraction studies[5] the $Nd^{3+}$ ions do not show a long range magnetic order until the lowest measured temperatures.

Figure 2 shows the low temperature dependence of the specific heat ($C$) with zero applied magnetic field for the $Nd_{0.5}Sr_{0.5}MnO_3$ and $Nd_{0.5}Ca_{0.5}MnO_3$ samples. Here, the most significant feature is the presence of an *S-shape* or *Schottky-like* anomaly at low temperatures for both compounds. No similar anomaly have been reported for the $La_{0.5}Ca_{0.5}MnO_3$ compound[7].

Although it is tenting to establish a straight forward correlation between the intrinsic magnetic moment of the $Nd^{3+}$ ions and the *Schottky-like* anomaly, a sample of $Pr_{0.63}Ca_{0.37}MnO_3$ disproved such a simple relation. $Pr^{3+}$ ions have an intrinsic magnetic moment similar to $Nd^{3+}$ ions. However, V. Hardy et. al.[8] did not found a *Schottky-like* anomaly in the low temperature specific heat of a $Pr_{0.63}Ca_{0.37}MnO_3$ single crystal.

The temperature for the maximum in the specific heat ($T_S$) was 2.73 and 5.08 K for the $Nd_{0.5}Sr_{0.5}MnO_3$ and $Nd_{0.5}Ca_{0.5}MnO_3$ samples, respectively. J. E. Gordon et. al.[9] found a similar anomaly in the specific heat, with the maximum at 4.44 K, for a $Nd_{0.67}Sr_{0.33}MnO_3$ polycrystalline sample. These values suggest an inverse correlation between the temperature of the peak (or alternatively, the energy splitting of the Schottky levels) and the effective ionic radio of the A-site in the perovskite compounds $AMnO_3$. The effective ionic radio of the A-site among these samples is greatest in $Nd_{0.5}Sr_{0.5}MnO_3$ and smallest in $Nd_{0.5}Ca_{0.5}MnO_3$.

It is also important to stress that the values of the specific heat at 2 K for $Nd_{0.5}Sr_{0.5}MnO_3$ and $Nd_{0.5}Ca_{0.5}MnO_3$ samples are more than 100 times higher than the corresponding values in $La_{0.67}Ba_{0.33}MnO_3$ and $La_{0.80}Ca_{0.20}MnO_3$, as was reported by J. J. Hamilton et. al.[10]. These enhanced excitation energies near the ground state could be partly due to the existence of a charge and orbital ordering in the first two compounds.

The continuous lines in figure 2 represent the best fit to the experimental data of the following expression:

$$C = C_S \cdot N_A \cdot k_B \frac{\left(\Delta_S/k_B T\right)^2 e^{\Delta_S/k_B T}}{\left(1+e^{\Delta_S/k_B T}\right)^2} + \gamma \cdot T + \sum_{n=1}^{4} \beta_{2 \cdot n+1} \cdot T^{2 \cdot n+1}$$

where the first term in the right side represents a two-level Schottky anomaly[11], the second one the electronic contribution and the third one represents the contribution associated to the lattice oscillations (or phonons). $N_A$ is the Avogadro´s number, $k_B$ is the Boltzmann´s constant, $C_S$ is the relative number of Schottky centers, $\Delta_S$ is the energy separation between the two Schottky levels, and $\gamma$ and $\beta_{2n+1}$ are coefficients characterizing the electronic and phononic contributions, respectively. Values of $\Delta_S$ were introduced as fixed parameters after being calculated from $\Delta_S = k_B T_S / 0.418$, an expression valid for the two-level Schottky equation[11].

Table 1 displays values of the physically relevant parameters. Although the two-level Schottky equation do not fit well the experimental data at low temperatures, the evaluated parameters are roughly in agreement to what is reported in the literature[7, 8, 9, 10]. Since all these samples are insulators at low temperatures, the contribution to the $\gamma$-term from free electrons is expected to be zero. However, we found a non zero value of $\gamma$, which may be associated to an extra contribution of the charge and orbital ordering. A more detailed study of the specific heat features is now in progress and will be published elsewhere.

Concluding, we presented low temperature specific heat and magnetization data for polycrystalline samples of $Nd_{0.5}Sr_{0.5}MnO_3$ and $Nd_{0.5}Ca_{0.5}MnO_3$. We could not found a straight forward correlation between the intrinsic magnetic moment of $Nd^{3+}$ ions and the Schottky-like anomaly. We thank the Brazilian science agencies FAPESP and CNPq for financial support.

[1] Gang Xiao, G. Q. Gong, C. L. Canedy, E. J. McNiff, Jr. and A. Gupta, J. Appl. Phys. 81 (8) 5324 (1997).

[2] Y. Moritomo, Phys. Rev. B 60 (14) 10374 (1999).

[3] J. López, P. N. Lisboa-Filho, W. A. C. Passos, W. A. Ortiz and F. M. Araujo-Moreira, Journal of Magnetism and Magnetic Materials 226-230, 507 (2001). Also at http://arxiv.org/abs/cond-mat/0004460

[4] J. López, P. N. Lisboa Filho, W. A. C. Passos, W. A. Ortiz, F. M. Araujo-Moreira, Kartik Ghosh, O. F. de Lima, and D. Schaniel, Phys. Rev. B 63 (22), 224422 (2001). Also at http://arxiv.org/abs/cond-mat/0103305.

[5] F. Millange, S. de Brion and G. Chouteau, Phys. Rev. B 62 (9) 5619 (2000).

[6] T. Vogt, A. K. Cheetham, R. Mahendiran, A. K. Raychaudhuri, R. Mahesh and C. N. R. Rao, Phys. Rev. B 54 (21) 15303 (1996).

[7] V. N. Smolyaninova, Amlan Biswas, X. Zhang, K. H. Kim, Bog-Gi Kim, S-W. Cheong and R. L. Greene, Phys. Rev. B 62 (10) R6093 (2000).

[8] V. Hardy, A. Wahl, C. Martin and Ch. Simon, Phys. Rev. B (63) 224403 (2001)

[9] J. E. Gordon, R. A. Fisher, Y. X. Jia, N. E. Phillips, S. F. Reklis, D. A. Wright and A. Zettl, Phys. Rev. B 59 (1) 127 (1999)

[10] J. J. Hamilton, E. L. Keatley, H. L. Ju, A. K. Raychaudhuri, V. N. Smolyaninova and R. L. Greene, Phys. Rev. B 54 (21) 14926 (1996)

[11] C. Kittel, "Introduction to Solid State Physics", fifth edition, pag. 454 (1976)

| Samples | Cs | $\Delta_S$ (meV) | $\gamma$ (mJ/mole-$K^2$) | $\beta_3$ (mJ/mole-$K^4$) |
|---|---|---|---|---|
| $La_{0.5}Ca_{0.5}MnO_3$[7] | - | - | 0 | 0.14 |
| $Nd_{0.5}Sr_{0.5}MnO_3$ | 0.39 | 0.56 | 1.3 | 0.26 |
| $Nd_{0.5}Ca_{0.5}MnO_3$ | 0.49 | 1.05 | 0.12 | 0.26 |

Table 1. Relative number of Schottky centers, energy separation between the two Schottky levels, and the coefficients characterizing the electronic and phononic contributions, from our measured samples and from V. N. Smolyaninova et. al.[7].

Figure 1. Magnetization measurements with a 5 T applied magnetic field in FCW conditions for the three polycrystalline samples studied. Magnetization is given in Bohr magnetons per manganese ion. Note the increase of magnetization at low temperatures for the samples with $Nd^{3+}$ ions.

Figure 2. Specific heat as a function of temperature in zero magnetic field for $Nd_{0.5}Sr_{0.5}MnO_3$ and $Nd_{0.5}Ca_{0.5}MnO_3$ samples. The continuous lines represent the best fitting of the experimental data by a two-level Schottky function, plus thermal and electronic contributions.

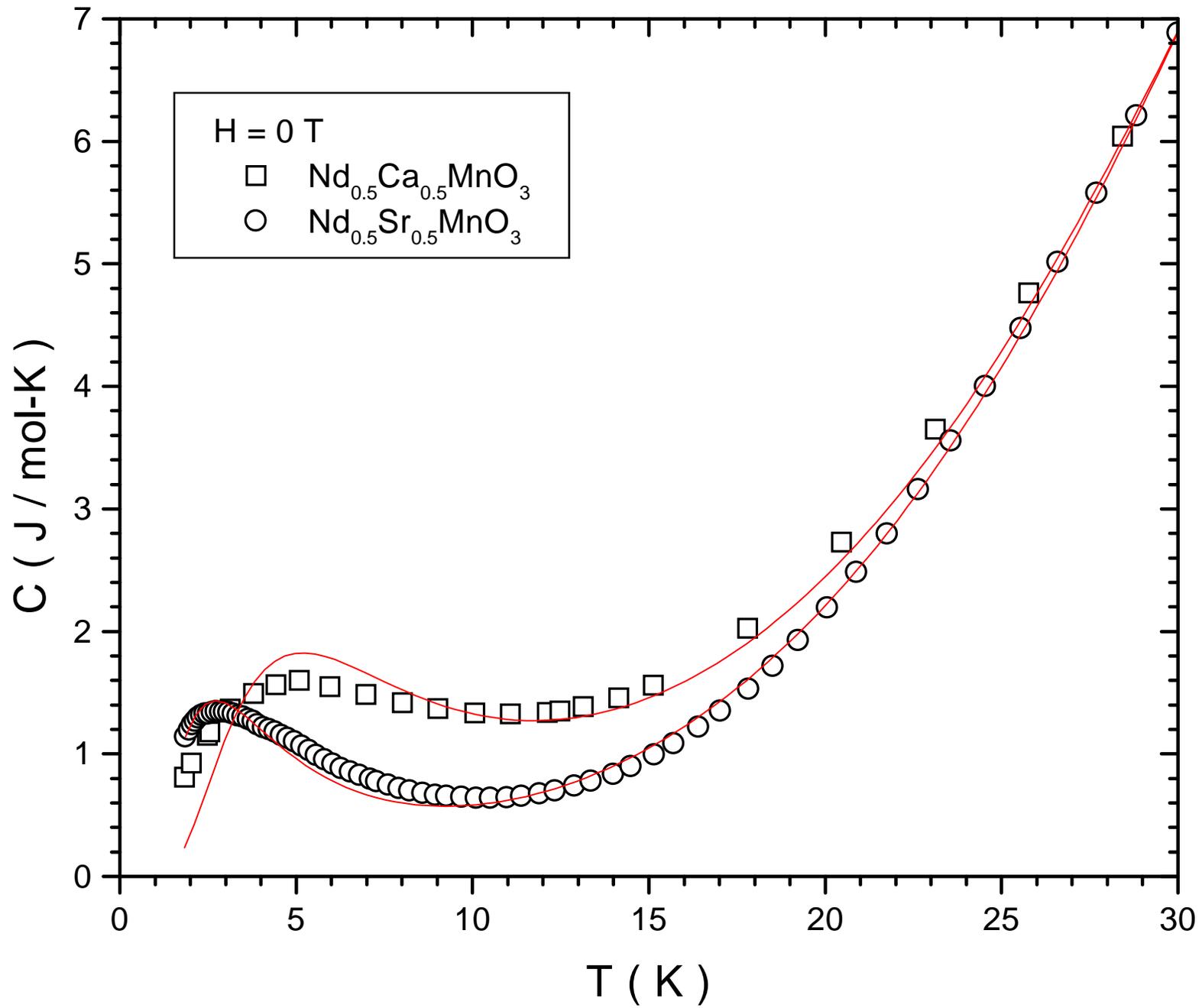

Figure 2